\documentclass[useAMS,usenatbib]{mn2e}
\usepackage{multirow}
\usepackage{times}
\usepackage{graphics,epsfig,txfonts}
\usepackage{textcomp}

\title[Discovery of SXP\,265]
  {Discovery of SXP\,265, a Be/X-ray binary pulsar in the Wing of the Small Magellanic Cloud
    \thanks{Based on observations with 
      XMM-Newton, an ESA Science Mission with instruments and contributions 
      directly funded by ESA Member states and the USA (NASA)}
  }
\author[R. Sturm et al.]{
        R.~Sturm,$^1$ 
        F.~Haberl,$^1$ 
        G.~Vasilopoulos,$^1$  
        E.~S.~Bartlett,$^2$  
        P.~Maggi,$^1$  
        A.~Rau,$^1$ \newauthor
        J.~Greiner$^1$ and
        A.~Udalski$^3$  \\
        $^1$Max-Planck-Institut f\"ur extraterrestrische Physik, Giessenbachstra{\ss}e, 85748 Garching, Germany\\
        $^2$Astrophysics, Cosmology and Gravity Centre (ACGC), Astronomy Department, University of Cape Town, Private Bag X3, Rondebosch, 7701, South Africa\\
        $^3$Warsaw University Observatory, Aleje Ujazdowskie 4, 00-478 Warsaw, Poland
      }

\date{Accepted 2014 August 14. Received 2014 August 14; in original form 2014 July 22}

\pagerange{\pageref{firstpage}--\pageref{lastpage}} \pubyear{2014}

\newcommand{\sxp}{SXP\,265}
\newcommand{\xmm}{{\it XMM-Newton}}
\newcommand{\swift}{{\it Swift}}
\newcommand{\ion}[2]{\mbox{#1\,{\sc #2}}}
\def\rahour{\hbox{\ensuremath{^{\rm h}}}}
\def\ramin{\hbox{\ensuremath{^{\rm m}}}}

\begin{document}

\label{firstpage}

\maketitle

\begin{abstract}
We identify a new candidate for a Be/X-ray binary in the \xmm\ slew survey and archival \swift\ observations
that is located in the transition region of the Wing of the Small Magellanic Cloud and the Magellanic Bridge.
We investigated and classified this source with follow-up \xmm\ and optical observations.
We model the X-ray spectra and search for periodicities and variability in the X-ray observations and the OGLE $I$-band light curve. 
The optical counterpart has been classified spectroscopically, with data obtained at the SAAO 1.9~m telescope, and photometrically, with data obtained using GROND at the MPG 2.2~m telescope.
The X-ray spectrum is typical of a high-mass X-ray binary with an accreting neutron star. 
We detect X-ray pulsations, which reveal a neutron-star spin period of $P_{\rm s} = (264.516\pm0.014)$~s.
The source likely shows a persistent X-ray luminosity of a few $10^{35}$~erg~s$^{-1}$ and in addition type-I outbursts that indicate an orbital period of $\sim$146~d.
A periodicity of 0.867~d, found in the optical light curve, can be explained by non-radial pulsations of the Be star. 
We identify the optical counterpart and classify it as a B1-2II-IVe star.
This confirms \sxp\ as a new Be/X-ray binary pulsar originating in the tidal structure between the Magellanic Clouds.
\end{abstract}

%% abstract without Latex: 
% We identify a new candidate for a Be/X-ray binary in the XMM-Newton slew survey and archival Swift observations that is located in the transition region of the Wing of the Small Magellanic Cloud and the Magellanic Bridge. We investigated and classified this source with follow-up XMM-Newton and optical observations. We model the X-ray spectra and search for periodicities and variability in the X-ray observations and the OGLE I-band light curve. The optical counterpart has been classified spectroscopically, with data obtained at the SAAO 1.9 m telescope, and photometrically, with data obtained using GROND at the MPG 2.2 m telescope. The X-ray spectrum is typical of a high-mass X-ray binary with an accreting neutron star. We detect X-ray pulsations, which reveal a neutron-star spin period of P = (264.516+-0.014) s. The source likely shows a persistent X-ray luminosity of a few 10^35 erg/s and in addition type-I outbursts that indicate an orbital period of ~146 d. A periodicity of 0.867 d, found in the optical light curve, can be explained by non-radial pulsations of the Be star. We identify the optical counterpart and classify it as a B1-2II-IVe star. This confirms SXP 265 as a new Be/X-ray binary pulsar originating in the tidal structure between the Magellanic Clouds.

\begin{keywords}
          galaxies: individual: Small Magellanic Cloud --
          galaxies: stellar content --
          stars: emission-line, Be -- 
          stars: neutron --
          X-rays: binaries.
\end{keywords}

\section{Introduction}

Besides supergiant high-mass X-ray binaries,
Be/X-ray binaries \citep[BeXRBs, for a review see ][]{2011Ap&SS.332....1R} are the dominant subclass of high-mass X-ray binaries (HMXBs).
They consist of a  Be star orbited by a compact object, usually a neutron star (NS).
Be stars primarily eject material in the equatorial plane, building up a decretion disc, 
which leads to observable emission lines (e.g. H$\alpha$) in the optical spectrum and an excess emission in the near infrared (NIR).
Both are potentially variable due to instabilities in the disc and interaction with the NS.
X-ray outbursts are observed when the NS accretes enhanced amount of material from this decretion disc,
particularly during periastron passage (type-I, $L_{\rm X}\ga$$10^{36}$~erg~s$^{-1}$) 
or decretion disc instabilities (type-II, $L_{\rm X}\ga$$10^{37}$~erg~s$^{-1}$), 
but persistent X-ray emission is also observed in some systems ($L_{\rm X}\sim$$10^{35}$~erg~s$^{-1}$).

Because of recent star formation, the Magellanic Clouds harbour a large population of BeXRBs that is well observable with today's X-ray observatories 
owing to the short distance of 50$-$60~kpc and a relatively small absorbing foreground column density of a few $10^{20}$~cm$^{-2}$.
These systems enable a wealth of possible physical studies.
Prominent examples are the recently discovered pulsar LXP\,169 \citep{2013A&A...554A...1M}, which shows optical transits likely caused by material captured by the NS,
and SXP\,1062 \citep{2012MNRAS.420L..13H,2012A&A...537L...1H,2013A&A...556A.139S}, which is associated to a supernova remnant allowing a robust age determination.
Both systems can be used to constrain accretion physics.
In addition to individually interesting sources, 
the known population of the Magellanic Clouds is as comprehensive as the Galactic sample and ideally suited for statistical studies,
e.g. to determine the NS spin distribution \citep[][]{2011Natur.479..372K,2014ApJ...786..128C}, 
the relation to the star-formation history \citep{2010ApJ...716L.140A},
and the faint end of the X-ray luminosity function \citep{2005MNRAS.362..879S,2013A&A...558A...3S}, and 
to find and constrain the population of Be/white dwarf systems \citep{2006A&A...458..285K,2012A&A...537A..76S,2012ApJ...761...99L,2013ApJ...779..118M}.

We have been granted two triggered \xmm\ observations in AO\,12 to follow up hard X-ray transients in the Magellanic Clouds and to get a more complete sample of pulsars in these galaxies.
The first source observed was RX\,J0520.5-6932 in the Large Magellanic Cloud (LMC)
leading to the discovery of the pulse period and a characterisation of the X-ray spectrum during a type-I outburst as presented by \citet{2014A&A...567A.129V}.
Here, we present the results from the second observation, that led to the discovery of a new pulsar, \sxp, 
located in the intersection of the Wing of the Small Magellanic Cloud (SMC) and the Magellanic Bridge, a tidal structure connecting the LMC and SMC
with a continuous stream of gas and young stars \citep{2014arXiv1405.7364S}.
This region has a different star-formation history \citep{2004AJ....127.1531H,2007ApJ...658..345H} and metallicity \citep{2008MNRAS.385.2261D}
to the Bar of the SMC, which harbours most of the known SMC BeXRBs,	
and so it is particularly interesting to find more BeXRBs in this region.

This paper is structured as follows:
In Section~\ref{sec:observations}, we describe the observations and reduction of the data,
followed by the analysis in Section~\ref{sec:analyses}, 
the discussion of the results in Section~\ref{sec:discussion},
and a summary in Section~\ref{sec:conclusion}.

\begin{table*}
 \centering
 \begin{minipage}{170mm}
  \caption{X-ray observations of \sxp.}
   \begin{tabular}{lcrrccrrrr}
  \hline
     \multicolumn{1}{c}{Observation} &
     \multicolumn{1}{c}{ObsID} &
     \multicolumn{1}{c}{Start time} &
     \multicolumn{1}{c}{End time} &
     \multicolumn{1}{c}{Instrument} &
     \multicolumn{1}{c}{Mode\footnote{Observation setup and filter: full-frame mode (ff) and photon-counting mode (pc).}} &
     \multicolumn{1}{l}{Net Exp} &  
     \multicolumn{1}{r}{Net Cts\footnote{Net counts as used for spectral analysis in the (0.2$-$12.0) keV band for \xmm\ and in the (0.3$-$7.0) keV band for \swift.}} &
     \multicolumn{1}{r}{R$_{\rm sc}$\footnote{Radius of the circular source extraction region.}}\\
     \multicolumn{1}{l}{} &
     \multicolumn{1}{l}{} &
     \multicolumn{1}{c}{(UT)} &
     \multicolumn{1}{c}{(UT)} &
     \multicolumn{1}{c}{} &
     \multicolumn{1}{l}{} &
     \multicolumn{1}{c}{[ks]} &
     \multicolumn{1}{c}{} &   
     \multicolumn{1}{r}{[\arcsec]}\\
 \hline
          XMM 2013  &  0724650301  &  2013 Oct 27 16:24  &  2013 Oct 28 01:46  &  EPIC-MOS1  &  ff--medium  & 24.0  & 3678    &  48  \\ 
                    &              &              16:24  &              01:46  &  EPIC-MOS2  &  ff--medium  & 24.0  & 4502    &  54  \\ 
                    &              &              16:47  &              01:42  &  EPIC-pn    &  ff--thin    & 20.2  & 11\,904 &  46  \\ 
     \noalign{\smallskip}\hline\noalign{\smallskip}
       Swift 2010 a &  00040893001 &  2010 Sep 22 01:10  &  2010 Sep 22 23:50  &  XRT        &  pc          & 6.6    & 52     &  40  \\  
       Swift 2010 b &  00040893002 &  2010 Oct 01 11:22  &  2010 Oct 01 16:33  &  XRT        &  pc          & 2.1    & 18     &  40  \\  
       Swift 2013 a &  00040893003 &  2013 Feb 16 16:46  &  2013 Feb 16 18:41  &  XRT        &  pc          & 2.0    & 41     &  40  \\  
       Swift 2013 b &  00040893004 &  2013 Oct 23 13:59  &  2013 Oct 23 15:56  &  XRT        &  pc          & 2.1    & 160    &  40  \\  
       Swift 2013 c &  00040893006 &  2013 Oct 31 10:58  &  2013 Oct 31 12:45  &  XRT        &  pc          & 2.1    & 65     &  40  \\  
\hline
\end{tabular}
\end{minipage}
\label{tab:xray-obs}
\end{table*}

\section{Observations and data reduction}
\label{sec:observations}

\begin{table*}
 \centering
 \begin{minipage}{170mm}
  \caption{Spectral-fit results for \sxp.}
  \begin{tabular}{llccccccccc}
  \hline
      \multicolumn{1}{l}{Observation} &
      \multicolumn{1}{l}{Model\footnote{For definition of spectral models see Section~\ref{sec:analyses:Xspec}.}} &
      \multicolumn{1}{c}{N$_{\rm H, SMC}$\footnote{Column density within the interstellar medium of the SMC or intrinsic to the source in $10^{21}$~cm$^{-2}$.}} &
      \multicolumn{1}{c}{$\Gamma$} &
      \multicolumn{1}{c}{$kT$\footnote{Temperature in keV.}} &
      \multicolumn{1}{c}{$R$\footnote{Radius of the emitting area (for BB) or inner-disc radius for an inclination of $\Theta=0$ (for DiskBB) in km.}} &
      \multicolumn{1}{c}{$F$\footnote{Observed flux in the (0.2$-$10.0)~keV band in $10^{-12}$~erg~cm$^{-2}$~s$^{-1}$. 
                                      For the \swift\ spectra, the best-fit power-law model derived from the \xmm\ data has been used to estimate the fluxes.}} &
      \multicolumn{1}{c}{$L_{\rm x}$\footnote{Source intrinsic X-ray luminosity in the (0.2$-$10.0)~keV band in $10^{36}$~erg~s$^{-1}$, corrected for absorption and assuming a distance of the source of 62.1~kpc.}} &
      \multicolumn{1}{c}{$\chi^2_{\nu}$} &
      \multicolumn{1}{c}{$\nu$} \\
  \hline
      XMM 2013  &  PL         &  $1.07_{-0.20}^{+0.21}$   & $0.820_{-0.021}^{+0.022}$ & --                 & --                         & $3.836\pm0.077$  & $1.83_{-0.03}^{+0.03}$ &  1.16 & 583 \\ \vspace{.7mm}
                &  DiskBB     &  $0.711_{-0.14}^{+0.15}$   & --                    & $6100_{-450}^{+540}$  & $0.101_{-0.012}^{+0.012}$     & $3.716\pm0.080$  &$1.77_{-0.03}^{+0.03}$  &  1.03 & 583 \\ \vspace{.7mm}
                &  PL+BB      & $4.58_{-0.88}^{+0.98}$    & $0.914_{-0.029}^{+0.034}$ & $63.7_{-5.8}^{+5.4}$  & $880_{-340}^{+610}$           & $3.786\pm0.076$  & $2.88_{-0.45}^{+0.32}$ &  1.07 & 581 \\ \vspace{.7mm}
                &  PL+BB\,2   & $0.493_{-0.17}^{+0.23}$    & $0.825_{-0.052}^{+0.055}$ & $1280_{-142}^{+142}$  & $1.20_{-0.16}^{+0.19}$        & $3.790\pm0.050$ & $1.70_{-0.02}^{+0.02}$  &  0.99 & 581 \\ \vspace{.7mm}
                &  PL+DiskBB  & $4.60_{-0.93}^{+0.95}$   & $0.926_{-0.030}^{+0.032}$ & $73.4_{-7.4}^{+7.0}$  & $800_{-320}^{+600}$            & $3.785\pm0.078$  & $3.37_{-0.69}^{+1.43}$ &  1.07 & 581 \\
      \noalign{\smallskip}\hline\noalign{\smallskip}\vspace{.7mm}
      Swift 2010 a  &  \multirow{2}{*}{PL}        & \multirow{2}{*}{$<$1.6}  & \multirow{2}{*}{$0.82\pm0.34$}             & \multirow{2}{*}{--}             &  \multirow{2}{*}{--}            & 
                                                                                             $0.69\pm0.17$ & $0.33_{-0.07}^{+0.08}$ &  \multirow{2}{*}{--} & \multirow{2}{*}{--} \\ \vspace{.7mm}
      Swift 2010 b  &            &                   &                     &                &                                & $0.89\pm0.41$ & $0.43_{-0.14}^{+0.19}$ &       &    \\ 
      \noalign{\smallskip}\hline\noalign{\smallskip}\vspace{.7mm}
      Swift 2013 a &             &            &                     &                &                                & $2.02\pm0.56$  & $0.96_{-0.21}^{+0.25}$ &       &     \\ \vspace{.7mm}
      Swift 2013 b &  PL         &   $<$3.4   & $0.75_{-0.20}^{+0.15}$ &  --            &   --                           & $7.06\pm0.96$ & $3.37_{-0.39}^{+0.43}$ &  --     &  --   \\ 
      Swift 2013 c &             &            &                     &                &                                & $2.87\pm0.63$ & $1.37_{-0.25}^{+0.28}$ &        &     \\ 
\hline
\end{tabular}
\end{minipage}
\label{tab:spectra}
\end{table*}

\subsection{XMM-Newton}
\label{sec:data:xmm}

\xmm\ \citep{2001A&A...365L...1J} carries three X-ray telescopes with the European Photon Imaging Camera \citep[EPIC, ][]{2001A&A...365L..27T,2001A&A...365L..18S} instrument at the focal point of each.
Between observations, the EPIC-pn operates with the medium filter whilst slewing, 
allowing for detections of sources with flux $F>1.2\times 10^{-12}$~erg~cm$^{-2}$~s$^{-1}$ in the (0.2$-$12.0)~keV band. 
These sources are included in the \xmm\ slew-survey catalogue \citep{2008A&A...480..611S}.

We found \sxp\ in an investigation of \xmm\ slew-survey detections in the field of the Magellanic Clouds.
The source is listed as XMMSL1\,J013250.6-742544 and was initially detected on 2007 June 4, followed by a second slew detection on 2007 October 28.
A third slew detection on 2011 October 17 is listed as XMMSL1\,J013251.0-742549 in the 1.6 release of the slew-survey catalogue.
Using a \swift\ observations on 2013 October 23 (see Section~\ref{sec:data:swift}), 
we found the source in a bright outburst allowing us to trigger a pointed \xmm\ observation performed 4 days later (MJD 56\,592.7$-$56\,593.1).
The observation log is presented in Table~\ref{tab:xray-obs}.

The data of the pointed observation of all three EPIC instruments were processed with SAS 13.0.0\footnote{Science Analysis Software (SAS), http://xmm.esa.int/sas/}. 
Time intervals of high background have been rejected by selecting background rates in the (7.0$-$15.0)~keV band below 8~cts~ks$^{-1}$~arcmin$^{-2}$ for EPIC-pn and below 2.5~cts~ks$^{-1}$~arcmin$^{-2}$ for both EPIC-MOS.
Events were extracted within a circular region around the source the radii of which was determined with the SAS task \textsc{eregionanalyse} to optimise the signal-to-noise ratio.
For the selection of background events, we used a circular extraction region of a source-free area on the same CCDs as the source.
For the creation of spectra and response matrices with \textsc{especget}, we used single- and double-pixel events of EPIC-pn and single- to quadruple-pixel events of EPIC-MOS with \textsc{FLAG=0}.
All spectra were binned for a signal-to-noise ratio of at least 5 in each bin.
Time series were extracted with the same pattern selection but using all events independent of flags.
The photon arrival times were randomised within the CCD frame time and recalculated for the solar-system barycentre.
A merged time series for all three instruments was created, using only simultaneous good-time intervals for all instruments.

\subsection{Swift}
\label{sec:data:swift}

\sxp\ was observed independently with the \swift\ satellite.
The source is listed as SWIFT\,J0132.5-7425, an unidentified X-ray source, in the \swift/BAT 58-month hard X-ray survey catalogue,
but is not included in the 70-month BAT catalogue \citep{2013ApJS..207...19B}.
SWIFT\,J0132.5-7425 was observed with two \swift/XRT exposures in 2010, also listed in Table~\ref{tab:xray-obs}.
The corresponding X-ray source in the 7-year Swift-XRT point-source catalogue \citep{2013A&A...551A.142D} is 1SWXRT\,J013251.3-742545.
We requested further \swift\ observations in 2013, after identifying the source as a BeXRB candidate.

\swift/XRT spectra were extracted from the cleaned level-3 event files with the \textsc{ftool}\footnote{http://heasarc.nasa.gov/ftools/} \textsc{xselect} 
using circular extraction regions for the source and background.
The spectra were not binned due to the low statistics. 
Ancillary response files were created using \textsc{xrtmkarf}.

\subsection{GROND}
\label{sec:data:grond}

On 2013 October 28 03:29 UT, close to the time of the \xmm\ observation, we observed \sxp\ 
with the Gamma-ray Burst Optical Near-ir Detector 
\citep[GROND, ][]{2008PASP..120..405G} at the MPG 2.2~m telescope in La Silla, Chile. 
141~s of integration were obtained in $g^{\prime}$, $r^{\prime}$, $i^{\prime}$, and $z^{\prime}$ and 240~s in $J$, $H$, and $K_{\rm S}$.
The data were reduced and analysed with the standard tools and methods described in \cite{2008ApJ...685..376K}. 
The $g^\prime$, $r^\prime$, $i^\prime$, and $z^\prime$ photometric calibration was obtained relative to an observation of an SDSS standard star field obtained $\sim$1 h earlier. 
The $J$, $H$, and $K_{\rm S}$ photometry was calibrated against selected 2MASS stars \citep{2006AJ....131.1163S}. 
The derived AB magnitudes including systematic uncertainties, but not corrected for foreground reddening, are: 
$g^{\prime}=14.87\pm0.02$,
$r^{\prime}=15.20 \pm0.03$,
$i^{\prime}=15.44 \pm0.04$,
$z^{\prime}=15.59 \pm0.07$,
$J =15.92 \pm0.07$,
$H =16.12 \pm0.08$, and
$K_{\rm S}=16.70 \pm0.11$~mag.

\subsection{OGLE}
\label{sec:data:ogle}

\sxp\ has been monitored regularly in the $I$ and $V$ band during the phase IV of the Optical Gravitational Lensing Experiment \citep[OGLE, ][]{2008AcA....58...69U} since 2010 August 6 (MJD 55\,414). 
The OGLE source identification is SMC739.11.1265.
In this study, we use data until 2014 January 12 (MJD 56\,669), containing 341 $I$-band and 29 $V$-band measurements collected in four seasons.
Ongoing observations are accessible with the X-Ray variables OGLE Monitoring \citep[XROM,][]{2008AcA....58..187U} system.

\subsection{SAAO 1.9~m}
\label{sec:data:optspec}

The optical spectra were taken with the 1.9~m Radcliffe telescope at the South African Astronomical Observatory (SAAO) on the night of 2013 November 5 (MJD 56\,601) with an exposure time of 1500~s. 
A 600~lines~mm$^{-1}$ reflection grating was used, blazed at 4600~\AA, along with the SITe CCD. 
A slit width of 1.5\arcsec\ was employed. This resulted in a wavelength range of $\lambda\lambda 3500-5500$~\AA\ and a resolution of $\sim3.0$~\AA, 
determined from the full width half maximum of the arc lines in the comparison spectra. 
This corresponds to 2.7 pixels on the CCD. The median signal-to-noise ratio in the $\lambda\lambda4000-5000$~\AA\ region is $\sim$$33$, ranging from $\sim$$17$ to 74.
 
The data were reduced using the standard packages available in the Image Reduction and Analysis Facility (\textsc{IRAF}). 
Wavelength calibration was implemented using comparison spectra of copper and argon lamps 
taken immediately before and after the observation with the same instrument configuration. 
The spectrum was normalised and a redshift correction applied corresponding to a recession velocity of the SMC of $158$~km~s$^{-1}$ \citep*{1987A&A...171...33R}.

\section{Analyses and Results}
\label{sec:analyses}

\subsection{X-ray spectrum}
\label{sec:analyses:Xspec}

\begin{figure}
  \resizebox{\hsize}{!}{\includegraphics[angle=-90,clip=]{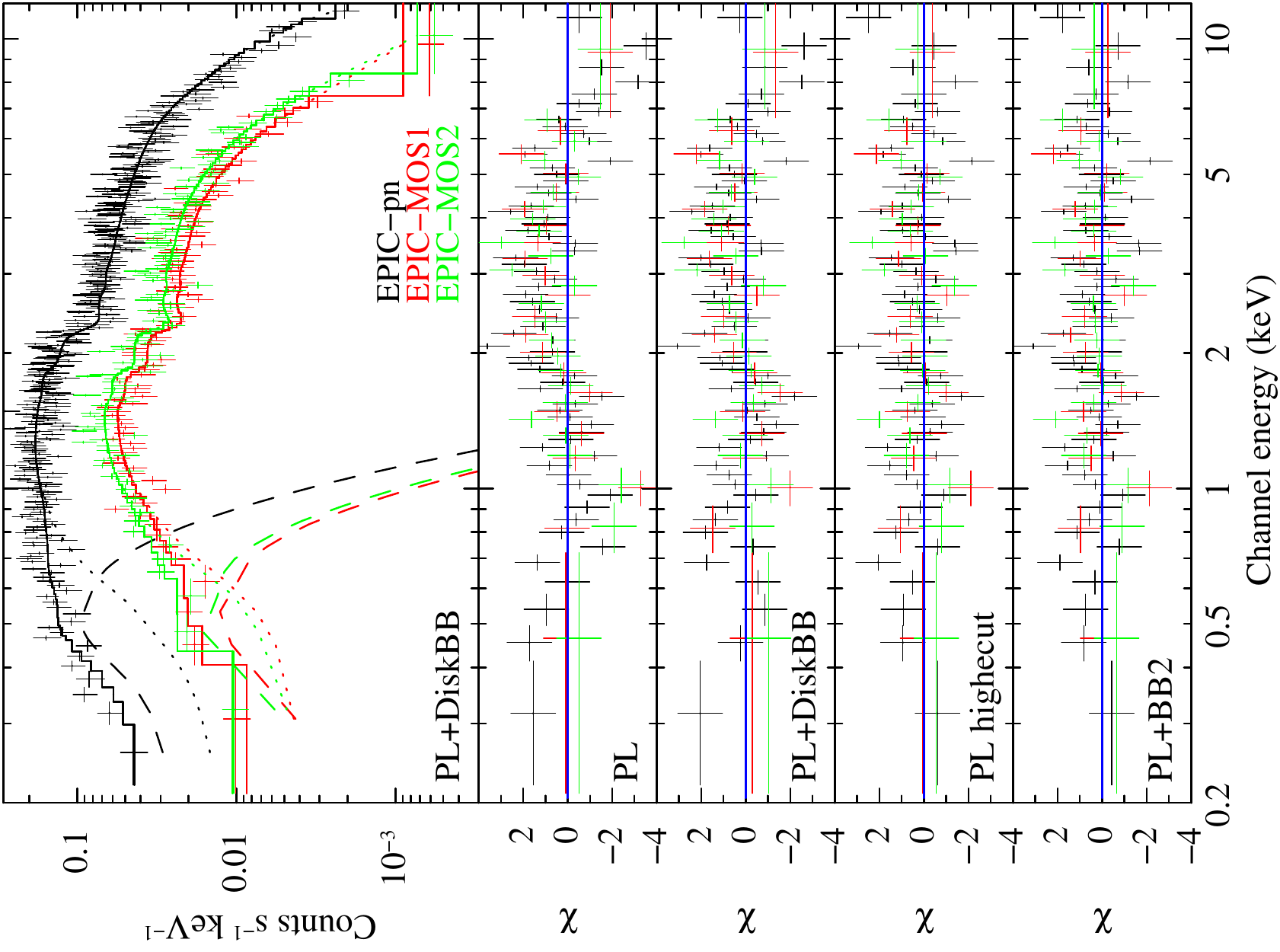}}
  \caption{ The X-ray spectra of \sxp\ as observed with \xmm\
            together with the folded best-fit model (solid lines) of a power law (dotted lines) and a multi-temperature disc (dashed lines)
            are plotted in the {\it upper panel}. 
            EPIC-pn/MOS1/MOS2 data are plotted in black/red/green.
            The residuals for a simple power-law, a power-law and disc, power-law with high-energy cut, and power-law and hot black-body model are plotted in the {\it lower panels} (from {\it top} to {\it bottom}).
            For clarity, the residuals have been rebinned by a factor of 5.
          }
  \label{fig:spec}
\end{figure}

The three EPIC spectra were fit simultaneously using \textsc{xspec} \citep{1996ASPC..101...17A} version 12.7.0.
A constant factor was included and allowed to vary for instrumental differences. 
We find $C_{\rm MOS1}=1.07\pm0.04$ and $C_{\rm MOS2} = 1.04\pm0.03$ relative to EPIC-pn ($C_{\rm pn}=1$), 
consistent with the current cross-calibration discrepancy.
All other parameters for various models were forced to be the same for all instruments and are listed in Table~\ref{tab:spectra} with 90 per cent confidence uncertainties ($\Delta \chi^2 = 2.71$).
The photoelectric absorption within the Galaxy was calculated 
assuming a column density of N$_{\rm H, Gal} = 4\times10^{20}$~cm$^{-2}$ \citep{1990ARA&A..28..215D}
with solar abundances \citep[according to][]{2000ApJ...542..914W}.
Additional absorption by material within the interstellar medium of the SMC or intrinsic to the source was determined by the fit with abundances set to  $Z=0.2Z_{\sun}$ \citep{1992ApJ...384..508R}.
The spectra are described satisfactorily by an absorbed power law with $\chi^2_{\nu}=1.15$ (with $\nu=583$ degrees of freedom),
but this fit exhibits systematic residuals in all instruments (second panel from top in Fig.~\ref{fig:spec}).

We found a disc black-body model to be a better fit to the data, but this requires physically implausible parameters.
Alternatively, we can account for a steeper spectral shape at higher energies with a broken power-law model, 
which results in $\chi^2_{\nu}=1.00$, $\Gamma_1=0.60$,  $\Gamma_2=1.1$, and a break energy of $E_{\rm br}=3.19$~keV,
or with a power-law model containing an additional high-energy cut off,
which results in $\chi^2_{\nu}=1.00$, $\Gamma=0.56\pm0.08$, a cut-off energy of $E_{\rm cut}=2.32\pm0.03$~keV, and a folding energy of $E_{\rm fold}=10.5_{-1.6}^{+2.4}$~keV.
We note, that a simultaneous fit to the \swift/BAT spectrum with this cut-off model (with free model normalisation, $C_{\rm BAT}=1.4\pm0.4$) 
does not show systematic offsets in the BAT data residuals. 
However, since this spectrum has only 8 data bins, the fit statistics is dominated by the \xmm\ spectra. 

A soft emission component \citep[e.g.][]{2004ApJ...614..881H,2008A&A...491..841E} might be expected to contribute to the X-ray emission of HMXBs: 
We tested a possible contribution by adding a black-body or a multi-temperature disc black-body model. 
These models give significantly better fits with f-test probabilities of $2\times10^{-8}$ and $1\times10^{-8}$. 
We found a second, even better, solution for a higher black-body temperature, as listed by the \textsc{PL+BB\,2} model in Table~\ref{tab:spectra}.

Another possible spectral feature of BeXRBs is Fe~K$\alpha$ line emission. By adding a Gaussian line profile with fixed central energy $E_{\rm c}=6.4$ keV and width of $\sigma=0$, 
we obtain a 3$\sigma$ upper limit for the line flux of $F_{\rm Fe K\alpha} \leq 2\times10^{-6}$~photons~cm$^{-2}$~s$^{-1}$,
which corresponds to an equivalent width of $EW\geq-54$~eV.

The low statistics of the \swift\ spectra mean the data can be described sufficiently with the best-fit power-law model derived from the \xmm\ data. 
When fit to the \swift\ spectra independently, the absorption and photon index are consistent with the \xmm\ values within uncertainties (see Table~\ref{tab:spectra}). 
We therefore used the parameter values from the best-fit power-law model and fitted only the normalisation to the unbinned spectra with C statistics, to derive the fluxes and luminosities listed in Table~\ref{tab:spectra}.

\begin{figure}
  \resizebox{\hsize}{!}{\includegraphics[angle=-90,clip=]{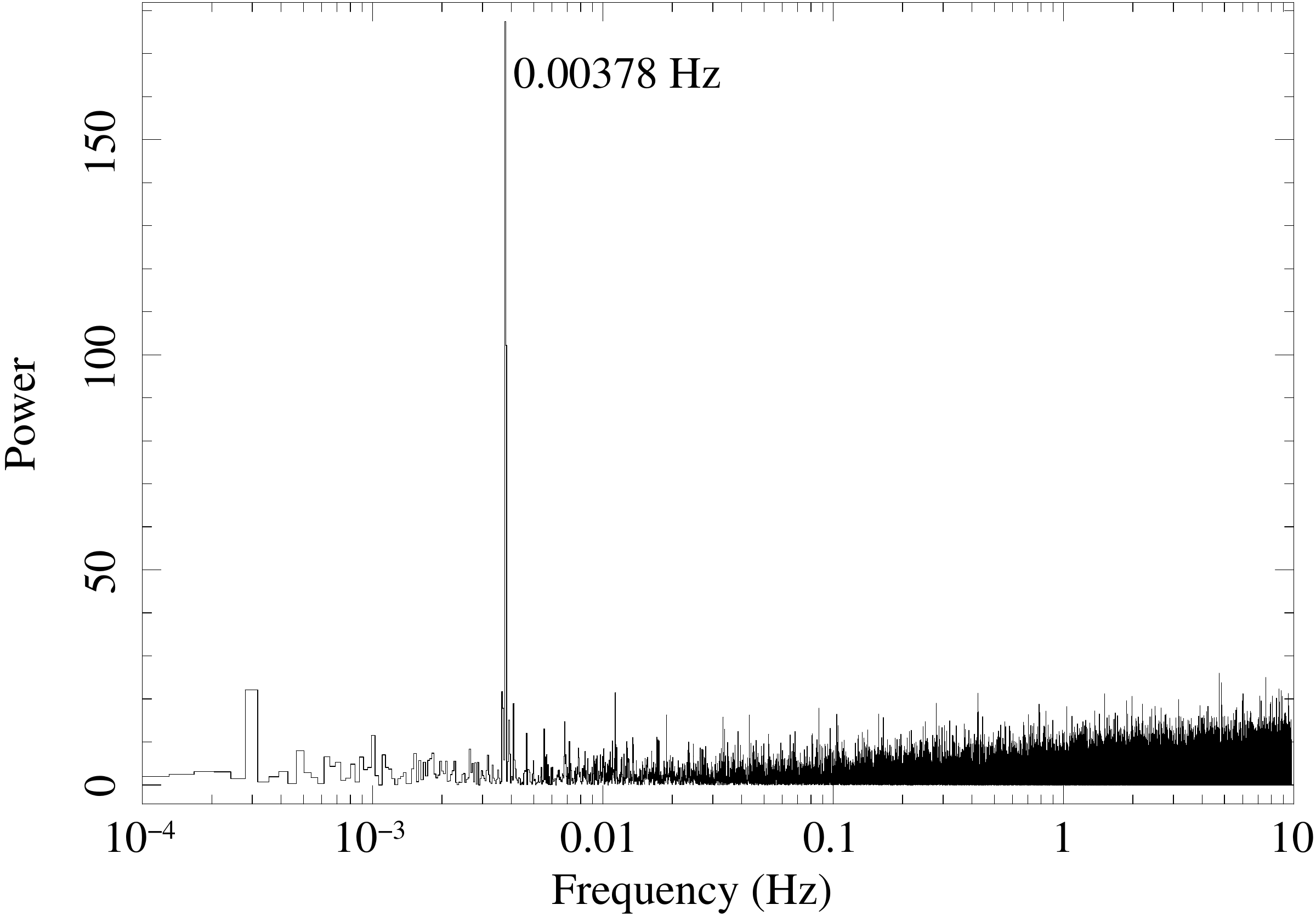}}
  \caption{Power-density spectrum of the merged EPIC time series in the (0.2$-$10.0)~keV band.}
  \label{fig:pds}
\end{figure}
\begin{figure}
  \resizebox{\hsize}{!}{\includegraphics[angle=0,clip=]{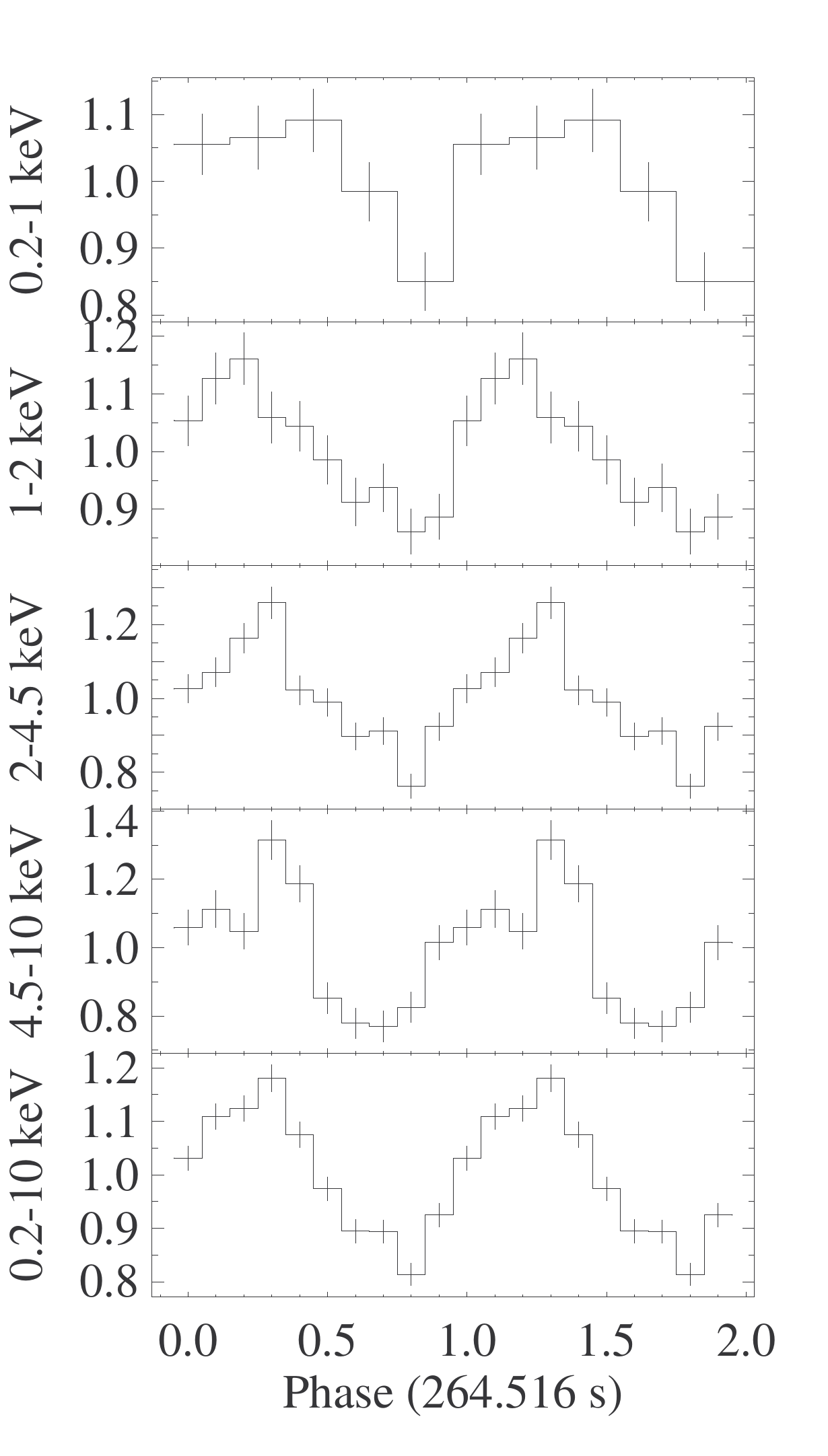}\includegraphics[angle=0,clip=]{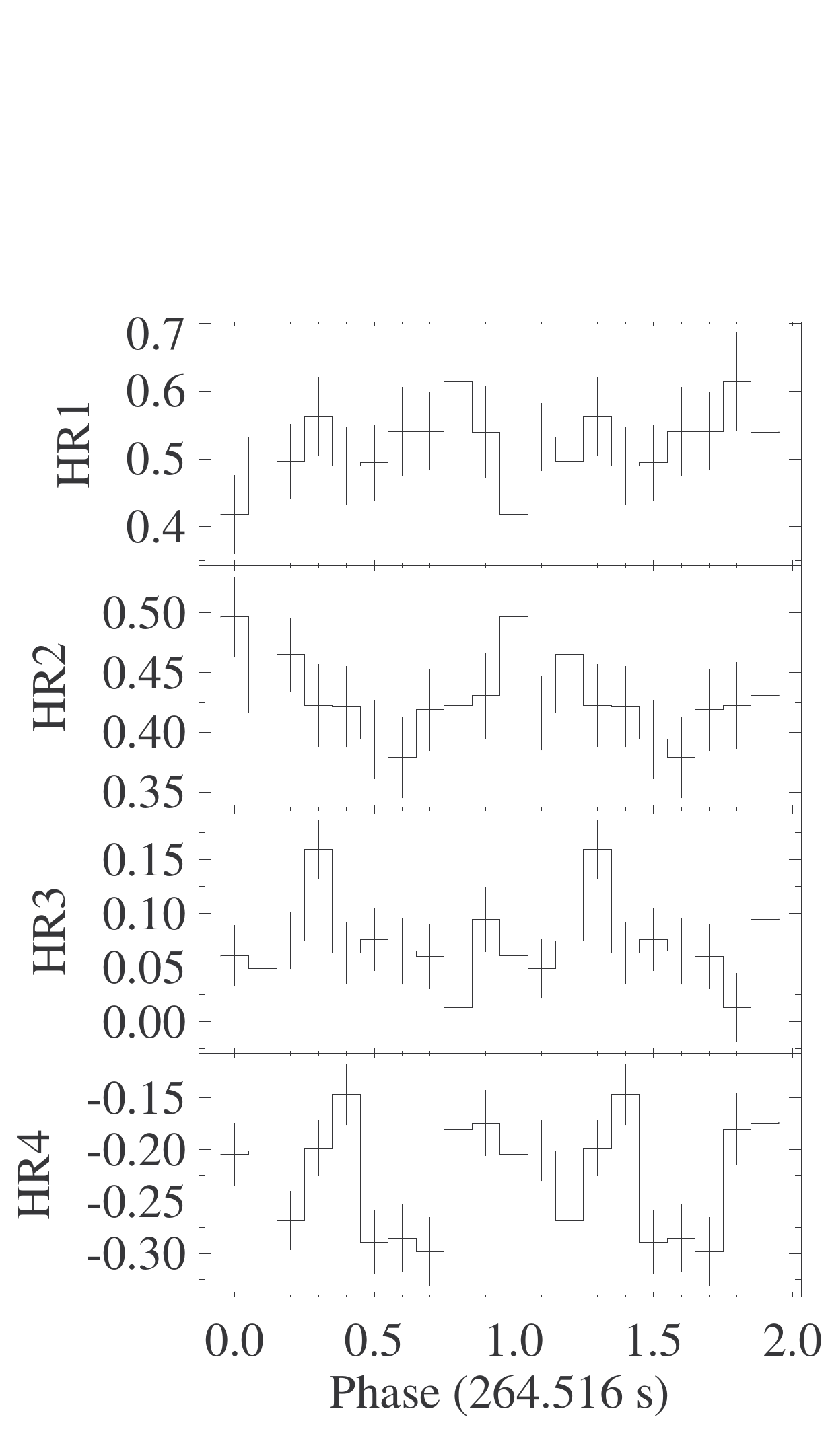}} 
  \caption{{\it Left:} X-ray pulse profile of \sxp\ in various energy bands of the merged time series.
           The pulse profiles are background-subtracted and normalised to the average net count rate of 
           0.14, 0.26, 0.30, 0.19 and 0.89~cts~s$^{-1}$ from {\it top} to {\it bottom}.
           {\it Right:} Hardness ratios as a function of pulse phase derived from the pulse profiles in two neighbouring standard energy bands.
          }
  \label{fig:pp}
\end{figure}

\subsection{X-ray pulsations}
\label{sec:analyses:Xpulse}

Using a fast Fourier transformation (FFT), we find a clear signal at a frequency of $f = 0.00378$~Hz in the merged EPIC time series in the (0.2$-$10.0)~keV band (Fig.~\ref{fig:pds}).
This period is independently seen in all three EPIC instruments.
The pulse period and its 1$\sigma$ uncertainty are determined as $P_{\rm s} = (264.516\pm0.014)$~s using a Bayesian detection method \citep[see][]{2008A&A...489..327H}.
The folded background-subtracted EPIC light curves are presented for the total (0.2$-$10.0)~keV band 
and the standard sub-bands (0.2$-$0.5), (0.5$-$1.0), (1.0$-$2.0), (2.0$-$4.5), and (4.5$-$10.0)~keV in Fig.~\ref{fig:pp}.
The first two bands have been merged to increase the statistics.
Variations in the hardness ratio, defined by $HR_{i}  = (R_{i+1} - R_{i})/(R_{i+1} + R_{i})$ with $R_{i}$ 
denoting the background-subtracted count rate in the standard energy band $i$ (with $i$ from 1 to 4), are also shown.
We estimate the pulsed fraction in the total energy band to be $R_{\rm pulsed}/R_{\rm total}=0.21\pm0.03$, assuming a sinusoidal pulse profile.

\subsection{Long-term X-ray light curve}
\label{sec:analyses:Xlcurve}

\sxp\ is listed three times in the \xmm\ slew survey catalogue \citep[][]{2008A&A...480..611S} 
as XMMSL1\,J013250.6-742544 and XMMSL1\,J013251.0-742549
with count rates  
$(1.52\pm0.57)$~s$^{-1}$ on 2007 June 4 (MJD 54\,255), 
$(0.70\pm0.25)$~s$^{-1}$ on 2007 October 28 (MJD 54\,401), and  
$(0.85\pm0.41)$~s$^{-1}$ on 2011 October 17 (MJD 55\,851), respectively.

Assuming the best-fit power-law model from above, this corresponds to (0.2$-$10.0)~keV fluxes of 
$(13.7\pm5.1)\times10^{-12}$,
$(6.3\pm2.3)\times10^{-12}$, and
$(7.7\pm3.7)\times10^{-12}$~erg~cm$^{-2}$~s$^{-1}$
for an EPIC-pn exposure with medium filter.

All X-ray fluxes from pointed observations are listed in Table~\ref{tab:spectra}.
These flux measurements reveal a variability by a factor of 10.
If the \xmm\ slew detections are included, variability of at least a factor of 20 becomes evident.

\begin{figure}
  \resizebox{\hsize}{!}{\includegraphics[angle=0,clip=]{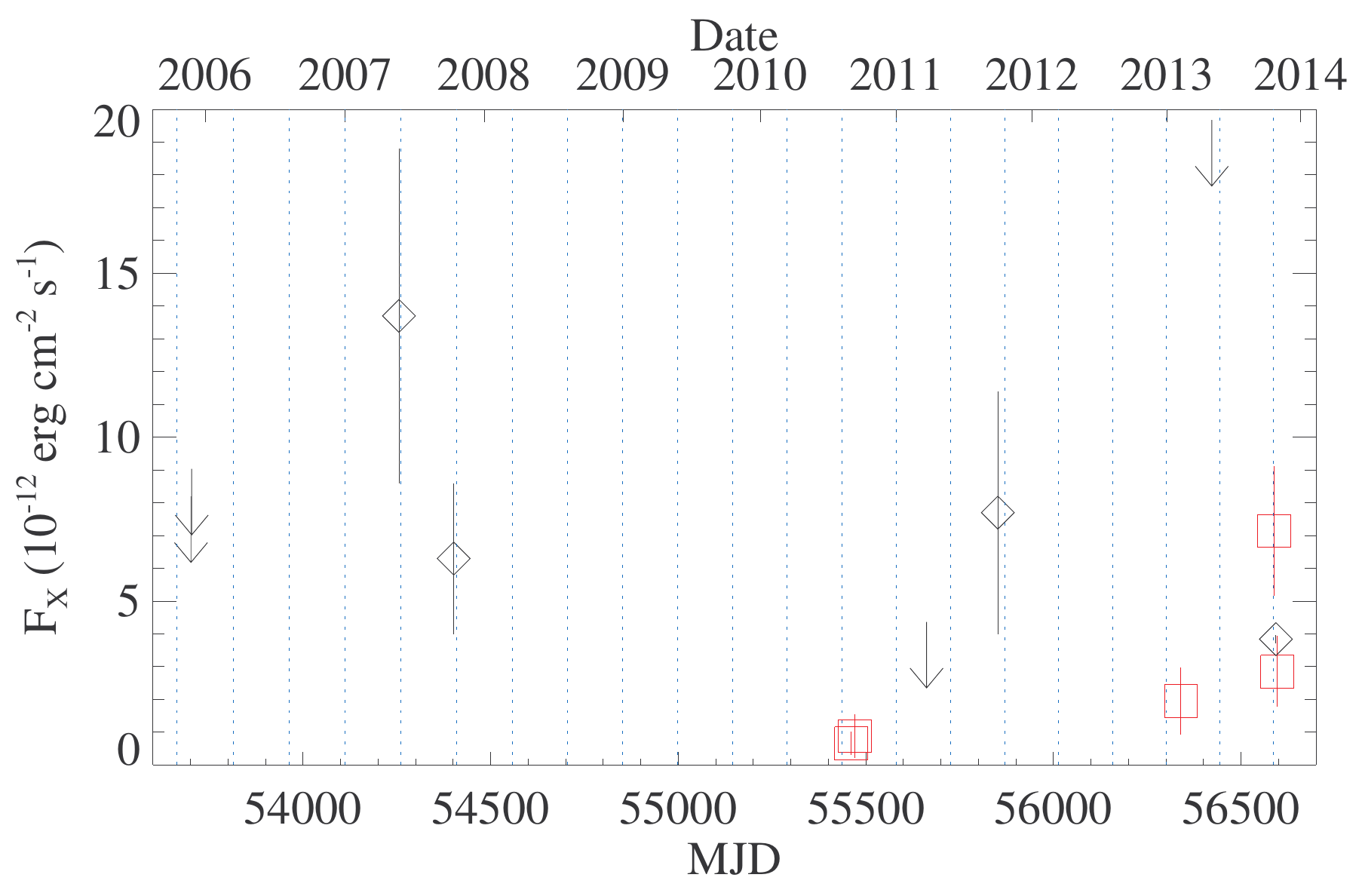}}
  \caption{ Long-term X-ray light curve including \xmm\ (black diamonds) and \swift\ (red squares) detections.
            Upper limits from the \xmm\ slew survey are marked by arrows.
            Dashed vertical lines indicate a 146~d period. 
          }
  \label{fig:lc_xray}
\end{figure}

The long-term X-ray light curve of \sxp\ is presented in Fig.~\ref{fig:lc_xray}.
Included are the upper limits from \xmm\ slews without a detection of \sxp\
that were obtained from the \xmm\ upper-limit server\footnote{
http://xmm.esac.esa.int/external/xmm\_products/slew\_survey/\newline upper\_limit/uls.shtml}.

Assuming a conservative detection limit of 7~cts in the ROSAT all sky survey,
we obtain an upper limit of 0.012~cts~s$^{-1}$ (0.1$-$2.0~keV) for 1990 October.
For the best-fit power-law model, this translates to a (0.2$-$10.0)~keV flux of 1.4$\times10^{-12}$~erg~cm$^{-2}$~s$^{-1}$
and thus does not exclude X-ray emission at the low-state level, as seen with \swift\ in 2010. 

By assuming a distance of the SMC of 62.1~kpc \citep{2014ApJ...780...59G} throughout the paper,
these fluxes translate to X-ray luminosities exceeding $10^{36}$~erg~s$^{-1}$ for the three \xmm\ slew detections and the outburst observed in 2013 October.
This would be atypical for a persistent X-ray emitting state and points to X-ray outbursts.
The first two slew detections are separated by 146~d and therefore unlikely to be caused by the same type-I outburst,
as these last typically up to $\sim$30 per cent of the orbit \citep{2008ApJS..177..189G}.
Interestingly, the third slew detection and the maximum of the 2013 outburst (\swift\ observation 2013 b), 
were 1451~d and 2187~d later, i.e. $\sim$10 times and $\sim$15 times the above separation.
This might indicate the orbital period of the system, but needs further confirmation.
We note that the X-ray light curved folded with the 146~d period 
includes all detections with $L_{\rm X} > 10^{36}$~erg~s$^{-1}$ 
within $\sim$10 per cent of the tentative orbital period
as expected for type-I outbursts \citep{2008ApJS..177..189G}.

\subsection{X-ray coordinates}
\label{sec:analyses:Xcoord}

The pointed \xmm\ observation provides the most precise X-ray position of \sxp.
Source detection was performed on the X-ray images of all three EPIC instruments simultaneously \citep[for details see][]{2013A&A...558A...3S}.
We identified seven other X-ray sources in the field of view, to derive an astrometric correction of 
$\Delta$RA $=-0.53$\arcsec\ and $\Delta$Dec $=-0.46$\arcsec.
This yields X-ray coordinates of \sxp\ of  
RA $= 01$\rahour32\ramin51\fs39 and 
Dec $= -$74\degr25\arcmin45\farcs4 (J2000).
Using astrometric corrections, the expected systematic uncertainty reduces from $\sigma_{\rm sys} = 1$\arcsec\ to $\sigma_{\rm sys} = 0.35$\arcsec\ \citep{2009A&A...493..339W}
resulting in a $1\sigma$ position uncertainty for \sxp\ of $\sigma = 0.36$\arcsec, where the statistical error is added quadratically.

This position is in agreement with the \xmm\ slew detections, 
with angular separations of 15.3\arcsec\ (due to the high position uncertainty this is equivalent to $1.7\sigma$), 3.9\arcsec ($0.66\sigma$), and  4.0\arcsec ($0.76\sigma$), respectively.
All \swift/XRT detections have angular separations $<$3.3\arcsec ($<$1.1$\sigma$) and are therefore also in agreement with the \xmm\ coordinates.

\begin{figure}
  \resizebox{\hsize}{!}{\includegraphics[angle=0,clip=]{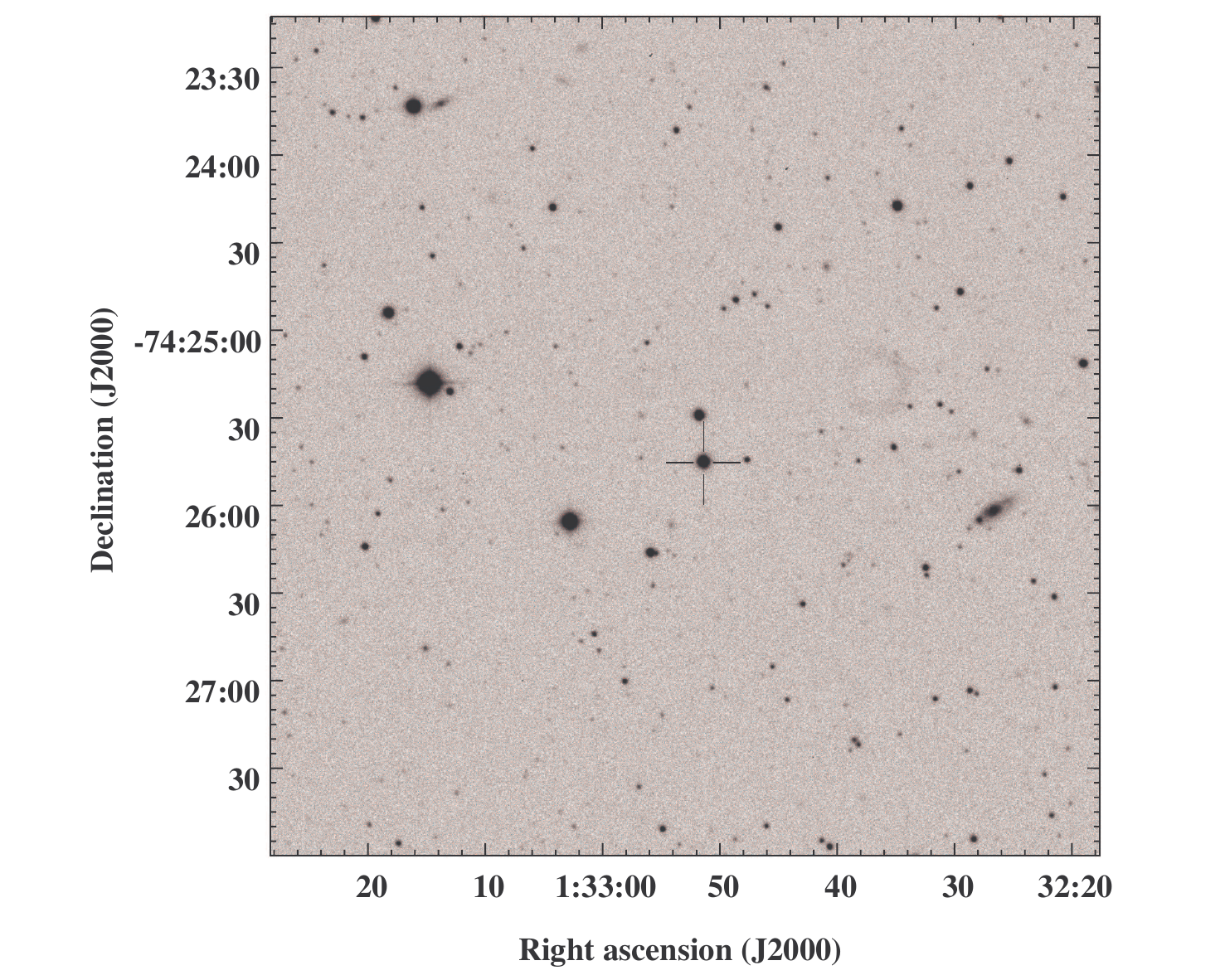}}
  \caption{ GROND $r'$-band finding chart. The cross marks the optical counterpart of \sxp. 
            In the 8\arcsec$\times$8\arcsec\ zoom-in, the corrected and uncorrected 1$\sigma$ \xmm\ positions are indicated by a small and large circle, respectively.
          }
  \label{fig:fc}
\end{figure}

The \xmm\ position allows us to identify 2MASS\,J01325144-7425453 as the optical counterpart with an angular separation of 0.26\arcsec.
The X-ray position is indicated in the GROND $r'$-band finding chart (Fig.~\ref{fig:fc}). 
The small circle indicates the improved and the large circle the uncorrected \xmm\ position. 
The star is also listed in the {\it Spitzer} SMC survey \citep{2011AJ....142..102G}
as SSTISAGEMA\,J013251.49-742545.2.
No other object is found to be in positional agreement with the X-ray source.

\subsection{Optical spectrum}
\label{sec:analyses:optspec}

\begin{figure*}
  \resizebox{\hsize}{!}{\includegraphics[angle=0,clip=]{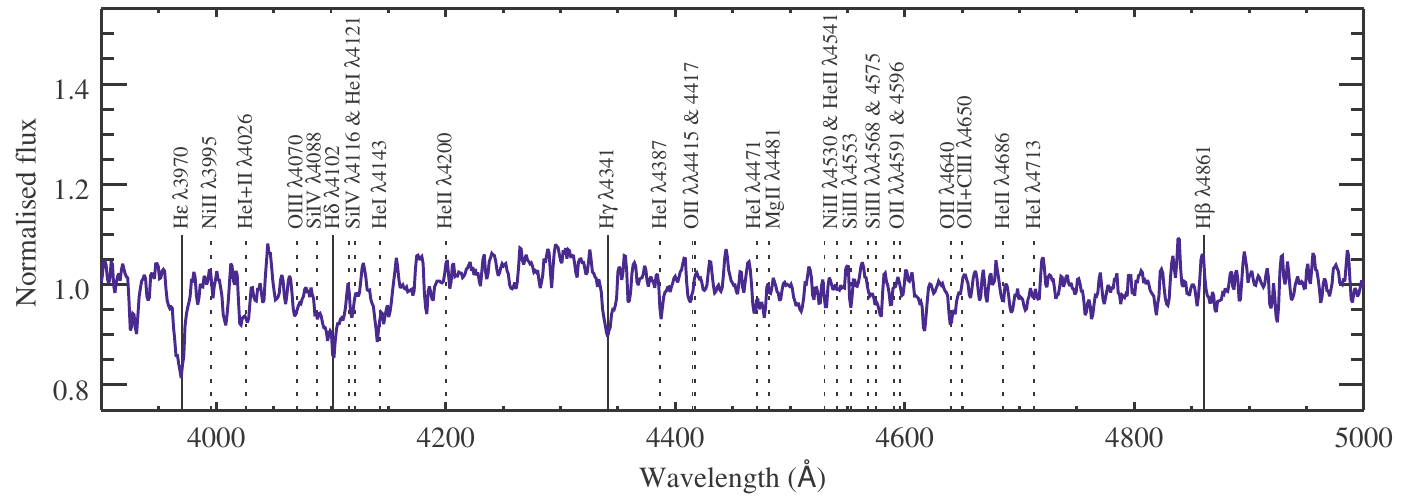}}
  \caption{Spectrum of \sxp\ in the wavelength range $\lambda\lambda$3900$-$5000~\AA\ taken with the Radcliffe 1.9~m telescope at SAAO on 2013 November 5. 
    The spectrum has been normalised and redshift corrected by $158$~km~s$^{-1}$. 
    Atomic transitions relevant to spectral classification have been marked.
  }
  \label{fig:optspec}
\end{figure*}

Spectrally classifying early-type stars in the SMC is difficult due to the low metallicity environment: 
The metal lines required for classification using the traditional Morgan-Keenan \citep[MK, ][]{1943QB881.M6.......} 
system are either much weaker or are not present. As such, the optical counterpart of \sxp\ was classified using the system developed by \citet{1997A&A...317..871L} for B-type supergiants in the SMC, 
and implemented for the SMC and LMC
by \citet{2004MNRAS.353..601E,2007A&A...464..289E}.
This system has been normalised to the MK system such that the classification criteria follow the same trends in line strengths.
Figure~\ref{fig:optspec} shows the normalised spectrum of \sxp\ smoothed with a boxcar average\footnote{http://northstar-www.dartmouth.edu/doc/idl/html\_6.2\newline/SMOOTH.html} 
with width 3. 

Be stars are characterised by their rapid rotation. This behaviour replenishes the decretion disc where the emission lines originate, which, in turn, leads to the `e' designation. 
It is therefore unsurprising that the optical spectrum of \sxp\ is dominated by the rotationally broadened hydrogen Balmer series. 
The H$\beta~\lambda4861$ line in particular shows evidence of `infilling' -- i.e. an emission feature superimposed on an absorption line.  

The spectrum does not show any evidence for the \ion{He}{ii} $\lambda\lambda$4200, 4541, or 4686 lines above the noise level of the data, suggesting a spectral type B1 or later. 
There does, however, appear to be some evidence for the \ion{Si}{iv} $\lambda\lambda$4088 and 4116 lines, which, if real, would constrain the spectral type to B1. 
The proximity of these lines to the broadened H$\gamma$ line makes it hard to determine whether they are genuine. 
The \ion{Si}{iii} $\lambda4553$ line is stronger than the \ion{Mg}{ii} $\lambda4481$ line constraining the spectral type to B2 or earlier.

The luminosity class of the counterpart was determined using the ratios of \ion{He}{i} $\lambda4121$/\ion{He}{i} $\lambda4143$ 
and \ion{Si}{iv} $\lambda4553$/\ion{He}{i} $\lambda4387$ \citep{1990PASP..102..379W}. 
The former decreases with increasing luminosity class (i.e. decreasing luminosity) 
the latter increases with increasing luminosity class. 
The signal-to-noise and resolution of the spectrum, along with its proximity to the Doppler broadened H$\delta$ line, make it difficult to draw any conclusions based on the \ion{He}{i} $\lambda4121$ line 
but the \ion{Si}{iv} $\lambda4553$/\ion{He}{i} $\lambda4387$ ratio suggests a luminosity class II-IV. This is supported by the strength of the \ion{O}{ii} spectrum, 
which also increases with increasing luminosity, however the low metallicity environment make this an unreliable luminosity-class indicator.

The well known distance to the SMC means we can calculate the absolute magnitude of the optical counterpart of \sxp\ accurately and precisely. 
This value can then be compared to those predicted for a B1-2II-IVe star to confirm the luminosity classification. 
Using the GROND $r'$ and $i'$ magnitudes, as well as equation 5 of \citet{2011A&A...526A.153K}, we get an uncorrected $V=(15.06\pm0.04)$~mag. 
For comparison, the OGLE $V$-band measurements cover the range from $V=14.986$ to $15.084$~mag  
and the one on 2013 November 4 (closest to the GROND observation) yielded $V=(15.040\pm0.003)$~mag.
Assuming the column density from the best-fit model to the X-ray spectrum of N$_{\rm H, SMC} = (5\pm2)\times10^{20}$~cm$^{-2}$, 
along with the N$_{\rm H, Gal}$ from \citet[][$4\times10^{20}$ cm$^{-2}$]{1990ARA&A..28..215D} 
and equation 1 from \citet{2009MNRAS.400.2050G}, we derive an optical extinction of $A_V=(0.407\pm0.092)$~mag. 
Along with a distance modulus of $\mu=(18.95\pm0.07)$~mag \citep{2014ApJ...780...59G}, 
this implies $M_V=(-4.3\pm0.1)$~mag. This value is consistent with a B1IIe star \citep{2006MNRAS.371..185W}, 
however we note that it falls within the range of a B1-1.5Ibe star all the way down to a B1-1.5IV-Ve star. 
As such, we classify the optical counterpart of \sxp\ as a B1-2II-IVe star.

\subsection{Spectral energy distribution}
\label{sec:analyses:sed}

\begin{figure}
  \resizebox{\hsize}{!}{\includegraphics[angle=0,clip=]{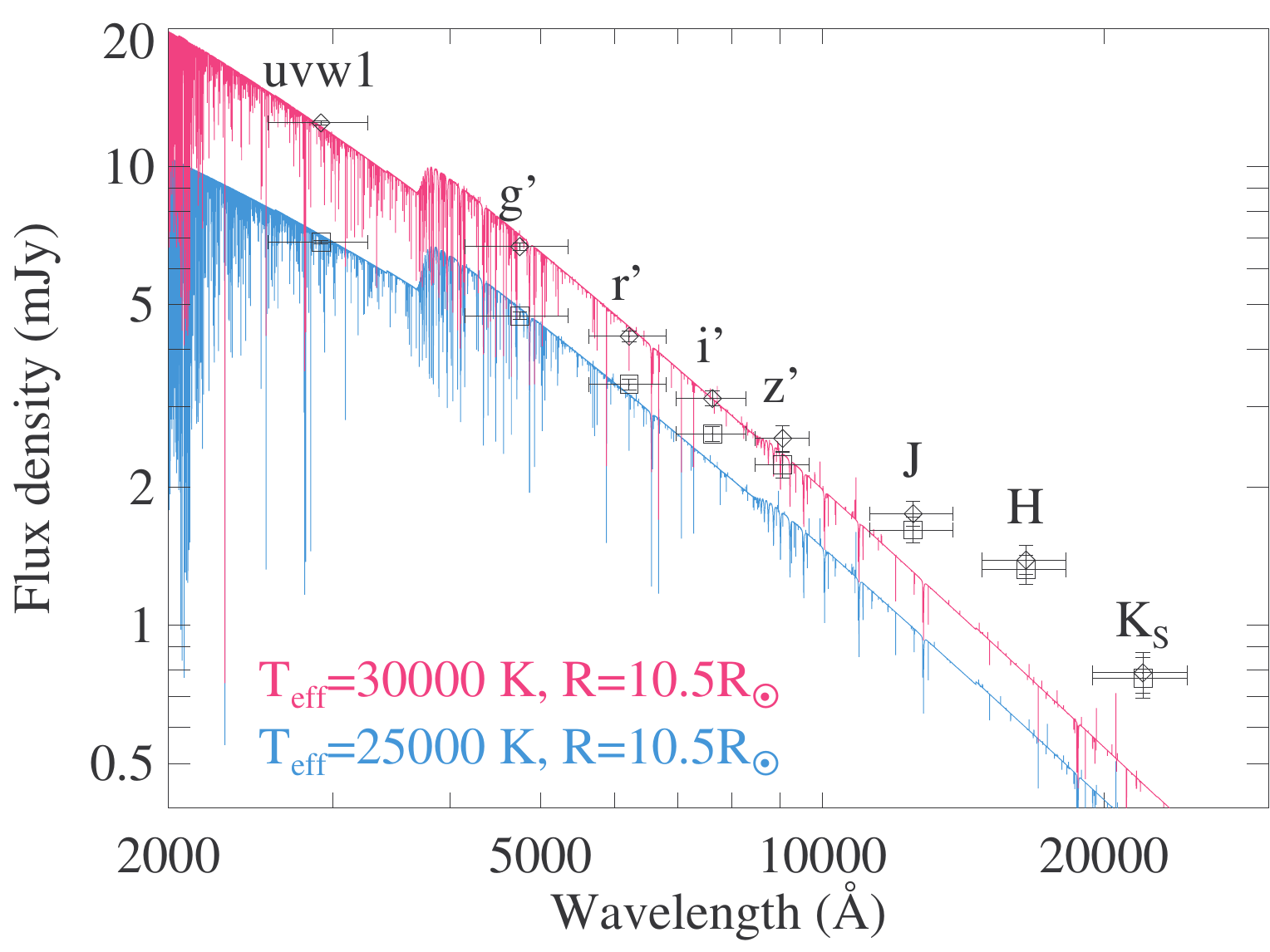}}
  \caption{Spectral energy distribution of the optical counterpart of \sxp\ corrected for Galactic foreground reddening (squares) and for additional maximum reddening within the SMC (diamonds).
           The red and blue lines represent stellar atmosphere models at $T_{\rm eff} =$ 25\,000~K and 30\,000~K, respectively, both for a stellar radius of $R=10.5R_{\sun}$.}
  \label{fig:sed}
\end{figure}

The simultaneously measured GROND data and the $uvw1$ measurement from the optical monitor of \xmm\ were used to construct the spectral energy distribution of the source. 
The boxes in Fig.~\ref{fig:sed} give flux densities, corrected for the Galactic foreground reddening of $E(B-V)=0.044$~mag \citep{2011ApJ...737..103S}, 
which we assume as the lower limit for the reddening. 
We  assume an upper limit of the reddening within the SMC of $E(B-V)=0.11$~mag, 
which corresponds to the total line-of-sight \ion{H}{i} column density of N$_{\rm H, SMC}=6\times 10^{20}$~cm$^{-2}$ \citep{1999MNRAS.302..417S} when using the relation of \citet{1995A&A...293..889P}.

We compare both spectral energy distributions with the stellar atmosphere models of \citet{2007ApJS..169...83L} with $Z=0.2Z_{\sun}$ and $\log{(g)}=4$.
The model with effective temperature $T_{\rm eff}=25\,000$~K and radius $R=10.5R_{\sun}$ (blue line) well describes the data at shorter wavelengths in the low-extinction case. 
This temperature is typical for a B1 star. 
The radius is somewhat larger than that expected for a main-sequence star.
In the high-extinction case, a $T_{\rm eff}=30\,000$~K (red line) is needed to compensate for the high extinction in the UV, this temperature is more typical of a B0 star.

The uncertainty in the proper extinction correction has the greatest effect towards the UV. 
However, towards the NIR, the reprocessing of the UV radiation from the star by the decretion disc is expected to causes an additional emission component.
For both extinction scenarios, we find a clear indication for such an excess in the $H$ and $K_{\rm S}$ band.

\subsection{Optical light curve}
\label{sec:analyses:opt-lc}

\begin{figure}
  \resizebox{\hsize}{!}{\includegraphics[angle=0,clip=]{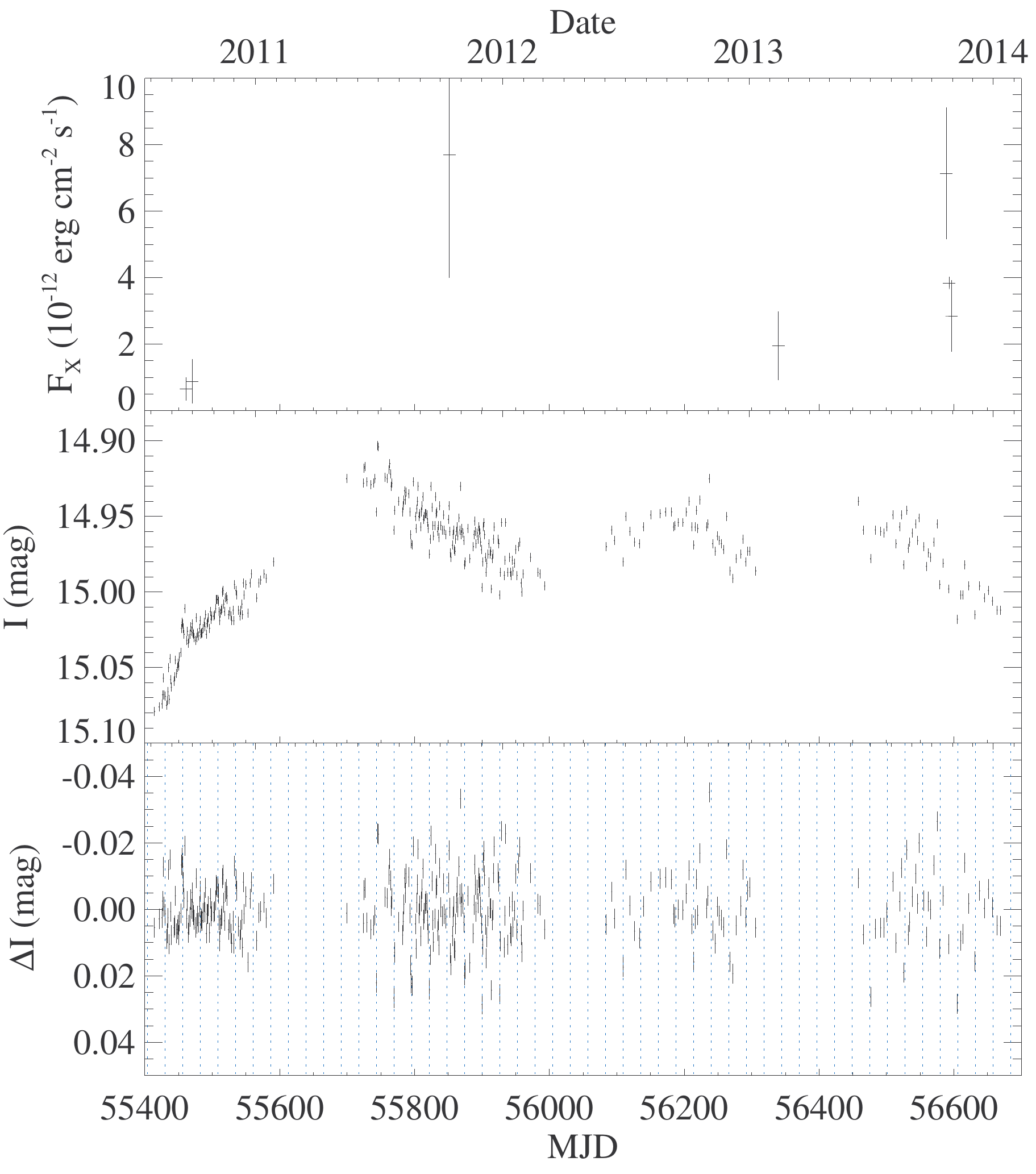}}
  \caption{ {\it Upper panel:} X-ray light curve in the (0.2$-$10.0)~keV band.
            {\it Middle panel:} OGLE $I$-band light curve.
            {\it Lower panel:} Detrended OGLE $I$-band light curve. Dashed vertical lines mark a 26.13~d period. 
          }
  \label{fig:lc}
\end{figure}

The OGLE-IV $I$-band light curve, presented in the middle panel of Fig.~\ref{fig:lc}, 
reveals long-term variability, e.g. by a systematic increase of $\sim$0.2~mag in the $I$ band between 2010 August and 2011 July (MJD 55\,414$-$55\,745).
Another, even stronger, brightening of the source is found between 2MASS (1998 October 8, MJD 51\,094) and 2MASS6X (2000 December 8, MJD 51\,886) observations \citep{2006AJ....131.1163S} 
by $(0.922\pm0.072)$~mag in the $J$ band, 
$(1.07\pm0.19)$~mag in the $H$ band, 
and $>0.5$~mag in the $K$ band.

\begin{figure}
  \resizebox{\hsize}{!}{\includegraphics[angle=0,clip=]{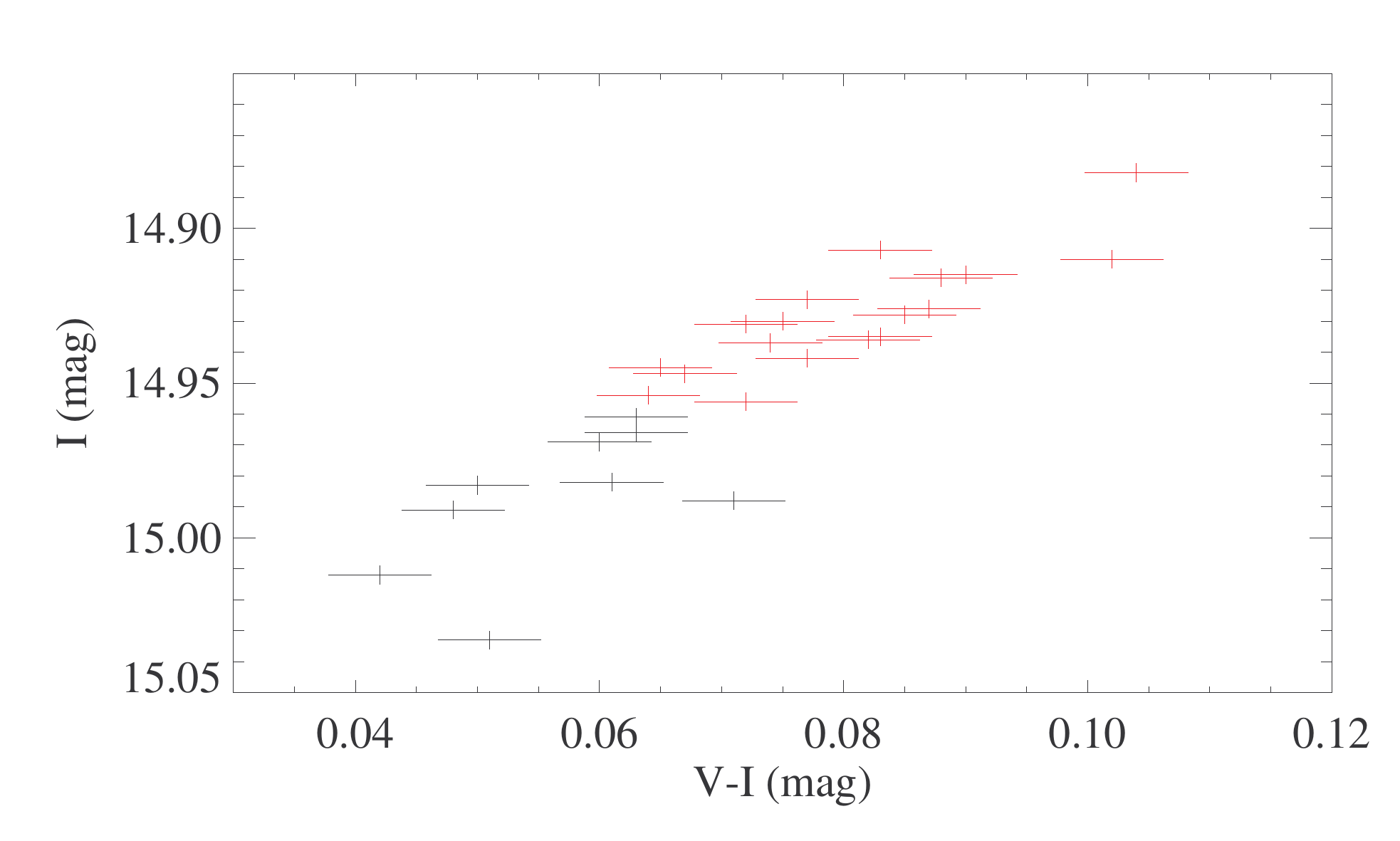}}
  \caption{ OGLE colour-magnitude diagram for \sxp\ from the first (black) and second (red) season.
          }
  \label{fig:cmd}
\end{figure}

We obtained $V-I$ colours using the OGLE $V$-band observations closest in time to the $I$-band observations, with a maximal separation of one day.
This resulted in 9 and 18 colour measurements in the first and second OGLE season, respectively,  as plotted in Fig.~\ref{fig:cmd}.
A clear correlation is seen between the parameters: The source becomes redder with increasing brightness.
This is expected if the inclination of the decretion disc is $\la80\degr$ with respect to the observer \citep*{2011MNRAS.413.1600R}.

\begin{figure}
  \resizebox{\hsize}{!}{\includegraphics[angle=0,clip=]{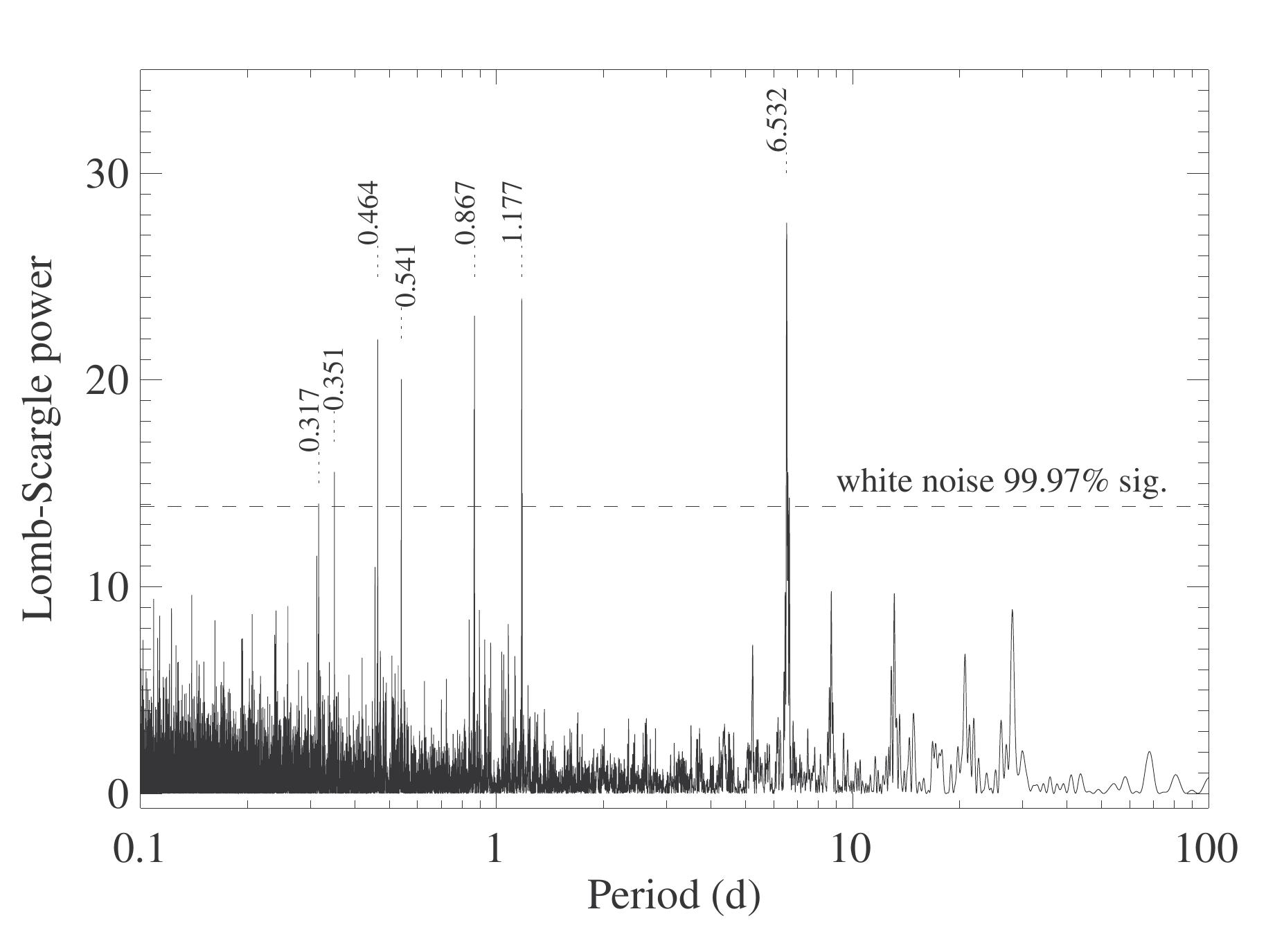}}
  \caption{Lomb-Scargle periodogram of the detrended OGLE light curve excluding the measurements before MJD 55\,600. The maxima of the strongest peaks are labelled.}
  \label{fig:ls}
\end{figure}

\begin{figure}
  \resizebox{\hsize}{!}{\includegraphics[angle=0,clip=]{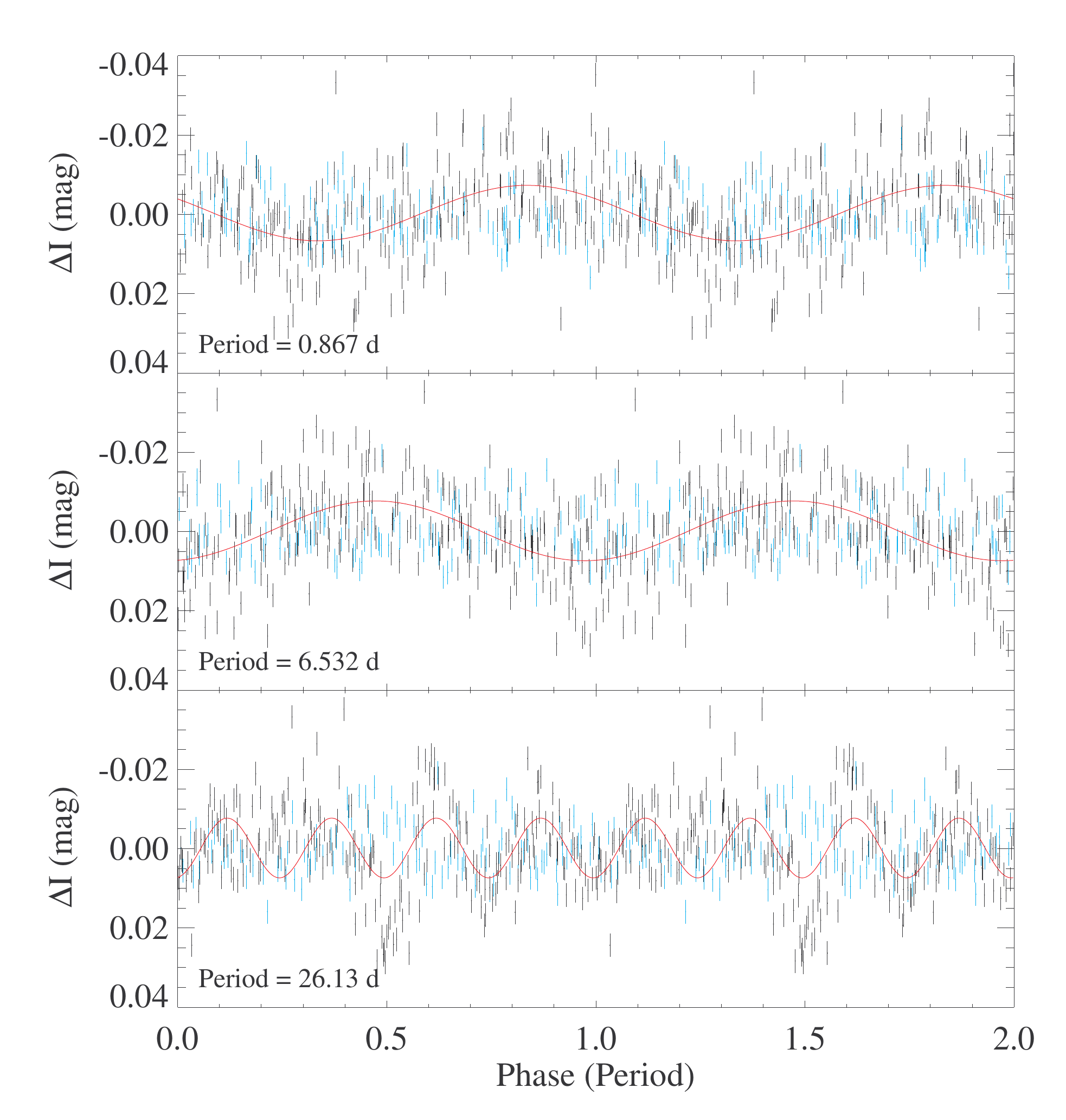}}
  \caption{Folded detrended OGLE light curve for several periods. The first season is plotted in cyan, others in black.
           The red line gives the best-fit sine function. In the lower panel, the sine function has a period of 6.532~d.}
  \label{fig:lc_con}
\end{figure}

The Lomb-Scargle periodogram \citep{1982ApJ...263..835S} of the detrended OGLE light curve, between 2011 May and 2014 January (season 2$-$4, MJD 55\,699$-$56\,669), is presented in Fig.~\ref{fig:ls}.
The significant peaks are found at 
6.532,
1.177,
0.867,
0.541,
0.464,
0.351, and
0.317~d
as labelled in Fig.~\ref{fig:ls}.
These are 1-day aliases of each other caused by the sampling.
Whereas the strongest power is found for the 6.532~d period when using the seasons 2$-$4,
the 0.867~d period has a similar power to the 6.532~d period, if only season 2 is used (22.00 vs. 22.02).
No significant detections are seen when the other seasons are independently investigated.
The folded light curves for both periods are sinusoidal (Fig.~\ref{fig:lc_con}) and have small amplitudes of 
$(0.00699\pm0.00028)$ and
$(0.00754\pm0.00028)$~mag, respectively.

To estimate the uncertainties in both periods, we use the bootstrap method \citep{1982jbor.book.....E}. We created random light curves from the original OGLE measurements 
by sampling with replacement (i.e. one epoch can be drawn multiple times) and searched for periodicities.
We repeated the above procedure 1000 times allowing the 1$\sigma$ uncertainties to be determined from the resulting distribution as
(6.532$\pm$0.012) and (0.867$\pm$0.010)~d.

In Fig.~\ref{fig:lc_con}, we also show the light curve convolved at 26.13~d, i.e. 4 times the 6.532~d period.
In this case, we see a stronger dip and increase at phase $\sim$0.5 than at the other expected minima.
These phases are also marked in the detrended light curve with vertical lines in Fig.~\ref{fig:lc}.
We note that these dips are not present in the first season, when the source was brightening but still in a fainter state.

The total OGLE light curve covers 1254 days allowing periods up to $\sim$600~d to be resolved.
The Lomb-Scargle periodogram of the unaltered light curve (i.e. without detrending) shows strong power at $\sim$241 and $\sim$643~d 
(in addition to the 1 and 6.53~d periods) and the $\sim$346~d that is seen when the first season is not used.
Due to the limited statistics and additional long-term variability, 
a longer coverage of the source is needed to establish the orbital period of the NS from optical variability.

\section{Discussion}
\label{sec:discussion}

The X-ray spectrum, X-ray pulsations, and the identification of the optical counterpart as a Be star 
allow us to clearly identify \sxp\ as a neutron-star BeXRB. Its properties are discussed in the following.

\subsection{Spectra}
\label{sec:discussion:spec}

\begin{figure*}
  \includegraphics[width=12cm]{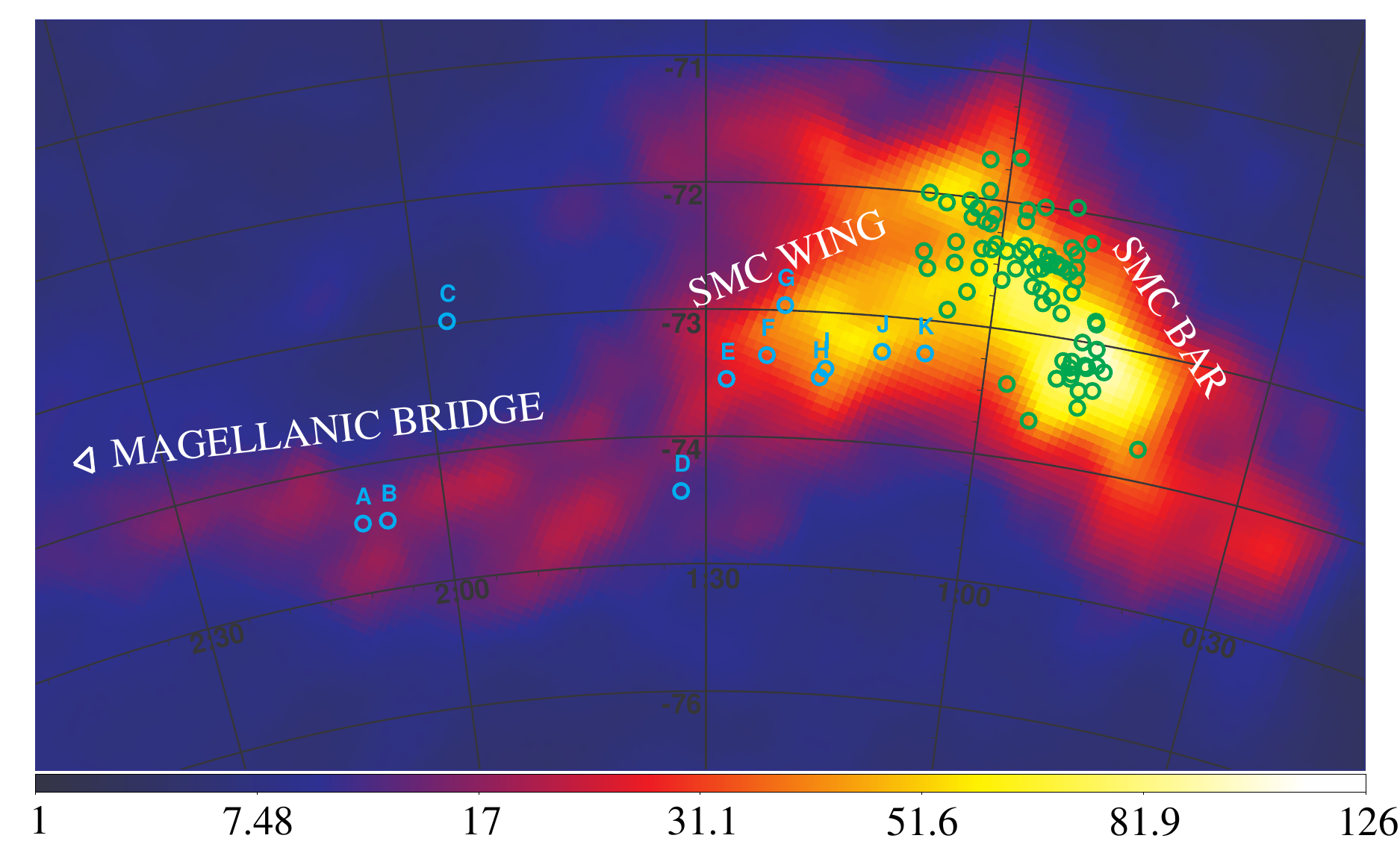}
  \caption{Spatial distribution of HMXB in the SMC compared to the \ion{H}{i} column density in $10^{20}$~cm$^{-2}$ from \citet{2009ApJS..181..398M}. BeXRBs and candidates in the SMC Bar are marked with green circles.
          HMXBs in the Wing and Bridge are plotted with cyan circles and labelled as follows:
          $^{\rm (A)}$ RXJ\,0209.6-7427 \citep{2005A&A...435....9K},
          $^{\rm (B)}$ SWIFT\,J0208.4-7428 \citep{2010MNRAS.403..709M},
          $^{\rm (C)}$ IGR\,J015712-7259 \citep{2010MNRAS.403..709M},
          $^{\rm (D)}$ SXP\,265 (this work),
          $^{\rm (E)}$ SXP\,1062 \citep{2012MNRAS.420L..13H},
          $^{\rm (F)}$ RX\,J0123.4-7321 \citep{2013A&A...551A..96S},
          $^{\rm (G)}$ IGR\,J01217-7257 \citep{2014ATel.5806....1C},
          $^{\rm (H)}$ SXP\,22.1 \citep{1997ApJ...474L.111C},
          $^{\rm (I)}$ SMC\,X-1,
          $^{\rm (J)}$ SXP\,31.0 \citep{1998IAUC.7062....1C},
          $^{\rm (K)}$ XMMU\,J010633.1-731543 \citep{2012MNRAS.424..282C}.
          }
  \label{fig:loc}
\end{figure*}

The precise \xmm\ coordinates allow us to clearly identify 2MASS\,J01325144-7425453 as the optical counterpart.
Our spectral classification of a B1-2II-IVe star is typical for a BeXRB in the SMC, 
as these are primarily found between O9.5 and B1.5 \citep{2008MNRAS.388.1198M}.
Irregular long-term optical variability and excess emission in the NIR are observed.
This is typical for Be stars and likely caused by a varying amount of reprocessing material in the decretion disc around the Be star.

The overall X-ray spectrum follows a power-law model with a typical photon index for BeXRBs in the SMC \citep{2004A&A...414..667H}. 
Evidence for a moderate deviation from this model is seen in the residuals. 
This could be due to the contribution of a low temperature ($kT \sim 100$~eV) soft excess. 
The inferred values are typical of those found in other BeXRBs \citep[e.g. ][]{2008A&A...484..451H,2011A&A...527A.131S} and 
allow for a soft excess with bolometric luminosity up to $L_{\rm bol}=1.6_{-0.8}^{+1.9}\times10^{36}$~erg~s$^{-1}$. 
However, these models require a high absorbing column. 
The derived radius of the emission region, estimated according to \citet{2004ApJ...614..881H}, is $R = ( L_{\rm X} / 4\pi\sigma T^4)^{1/2} = 756$~km; 
too large for an NS. We attribute this excess to an accretion disc or X-ray reprocessing material around the NS.

A contribution by a thermal component with higher temperature ($kT > 1$~keV) is also possible \citep[][]{2013MmSAI..84..626L,2013MNRAS.436.2054B}. 
For some sources this solution results in a physically questionable scenario, 
with the power-law component dominating the spectrum at lower energies and the black-body component contributing most of emission at higher energies. 
In this case, however, the black-body component contributes 20 per cent to the measured (0.2$-$10.0)~keV flux and the flux density of the black-body is below that of the power-law component at all energies.
A possible interpretation for the origin of this component is emission from the heated polar caps of the NS. 
The radius is at the upper limit of what is usually reported, but a slowly spinning NS with a rather constant X-ray luminosity is in agreement with this picture. 
An indication of spectral variability at higher energies is seen in the $HR$ variations (Fig.~\ref{fig:pp}), 
but the statistics is insufficient for a detailed phase-resolved spectral analysis. 

We also derived good fits using a power-law model with a high-energy cut off.  
In other HMXBs such a cut off is observed at higher energies, $\ga$10~keV \citep{2011MNRAS.410.1813T}, and so we do not favour this model.

For all the models, the absorption of the X-ray spectrum is in agreement with or above the total line-of-sight SMC column density of N$_{\rm H, SMC}=6\times 10^{20}$~cm$^{-2}$ \citep{1999MNRAS.302..417S}.
This suggests the system is behind the absorbing interstellar medium in the SMC or there is some intrinsic absorbing material in close proximity to the NS. 
Comparing the spectrum and the spectral energy distribution of the optical counterpart suggests a rather low extinction, placing the absorbing material close to the NS.

\subsection{Periodicities}
\label{sec:discussion:period}

The X-ray pulsations establish the compact object as an NS.
The spin period of $P_{\rm s} = (264.516\pm0.014)$~s puts the system in the population of slowly rotating  pulsars \citep{2011Natur.479..372K},
which typically show wide and circular orbits with only moderate X-ray outbursts \citep[][]{2014ApJ...786..128C}.
\sxp\ was consistently detected above $F_{\rm X}\sim10^{-12}$~erg~cm$^{-2}$~s$^{-1}$ when observed with sufficient sensitivity,
suggesting the source exhibits a rather persistent X-ray emission at this lower level.
We also observe what appears to be moderate type-I outbursts, where the X-ray luminosity increases by a factor of $\sim$10$-$20; these are only expected during periastron.
The temporal separation of these outbursts suggests an orbital period of $P_{\rm o}\sim(146\pm2)$~d, however this needs further confirmation. 
We note that this orbital period is in agreement with the spin period, according to the Corbet relation \citep{1984A&A...141...91C,2005ApJS..161...96L,2009IAUS..256..361C}.

A  6.53~d period is found in the OGLE light curve with a small amplitude of $(0.00754\pm0.00028)$~mag. 
There is also an indication of a feature at four times this period.
Given the long spin period of the NS, the orbital period is expected to be longer ($\ga50$~d) from the Corbet relation
and so we do not attribute this period to the orbit of the NS.
Some BeXRBs show short-term low-amplitude variability, 
which is believed to be caused by non-radial pulsations (NRPs) of the Be star. 
These periods are usually $<2$~d \citep*{2013MNRAS.431..252S}.
This seems an unlikely explanation for the 6.53~d period, but these NRPs may be responsible for the 0.867~d period (in which case the 6.53~d period would be artificial). 
If the 0.867~d period is slightly variable with time, its power in the periodogram might decrease with respect to aliases at longer periods \citep{2013MNRAS.431..252S}. 
Due to the lower statistics in the latter two OGLE seasons, we cannot test the variability of the period.

\subsection{Population comparison}
\label{sec:discussion:population}

The location of \sxp\ is compared with the \ion{H}{i} column density and the distribution of the other HMXBs in Fig.~\ref{fig:loc}.
\sxp\ is located in the transition region between the SMC Wing and the Magellanic Bridge.
The bulk of known HMXBs are found in the SMC Bar, which is known to have had an enhancement in its recent star formation.
A population of sources following the tidal feature towards the LMC, i.e. the Wing and high-density western part of the Bridge, is also evident. 
The HMXB population appears to be less dense in the Wing \citep{2008MNRAS.383..330M,2013A&A...558A...3S}, but 
the low X-ray coverage in the outer regions of the SMC Wing and the Bridge makes it difficult to currently estimate the population of HMXB. 
Only three confirmed BeXRBs are currently known to be located in the Magellanic Bridge, 
the first was found with ROSAT by \citet{2005A&A...435....9K} 
and two further systems were reported by \citet{2010MNRAS.403..709M}.

It is worth noting that no BeXRBs have been found elsewhere around the SMC, despite the similar coverage of the INTEGRAL and \xmm\ slew surveys. 
This suggests that the BeXRBs in the Magellanic Bridge do not originate in the SMC Bar, as noted by \citet{2010MNRAS.403..709M}.
The Bridge formed $\sim$200~Myr ago, i.e. much longer than the life time of an HMXB, 
and so these BeXRBs cannot have been tidally stripped from the SMC Bar population. 
The tidally triggered episode of star formation in the Bridge ended $\sim$40~Myr ago \citep{2007ApJ...658..345H}.
This is the expected evolution time of BeXRBs \citep{2010ApJ...716L.140A}, 
and it is therefore likely that the observed BeXRBs in the Bridge were formed in this event.

Future eROSITA survey observations \citep{2012arXiv1209.3114M} will reveal the population in the outer regions of the SMC with a sensitivity down to $L_{\rm X} \ga 10^{35}$~erg~s$^{-1}$.
This will allow us to further study the population of BeXRBs in a tidal structure,
and might put further constraints on supernova kick velocities when the surrounding regions of the SMC have a deeper homogeneous X-ray coverage.

\section{Summary and Conclusions}
\label{sec:conclusion}

We discovered a variable X-ray source, named \sxp, in archival \xmm\ and \swift\ observations. 
We classified this source as a BeXRB candidate based on its correlation with a blue star in the SMC
and investigated it in detail with additional follow-up observations with \xmm, \swift, GROND at the MPG 2.2~m telescope, 
and spectroscopy at the 1.9~m Radcliffe telescope at the SAAO in addition to the analysis of OGLE light-curve data.

The X-ray spectrum is typical for a HMXB with an NS compact object. 
The spin period of the NS is $P_{\rm s} = (264.516\pm0.014)$~s.
The source appears to show persistent X-ray luminosity at a few $10^{35}$~erg~s$^{-1}$ 
as well as type-I outbursts, of luminosity of a few $10^{36}$~erg~s$^{-1}$, indicating a possible orbital period of 146~d.

We identify the optical counterpart at 
RA = 01\rahour32\ramin51\fs47 and 
Dec = $-$74\degr25\arcmin45\farcs2 (J2000, 2MASS6X coordinates)
and classify it as a B1-2II-IVe star, which has long-term variability and an excess in the NIR.
An optical period is found at 0.867~d (or one of its 1-day alias) and might be explained by non-radial pulsations of the Be star.
\sxp\ is located in the transition region of the SMC Wing and the Magellanic Bridge 
where only a few systems are known and is the second most eastern pulsar associated with the SMC.

\section*{Acknowledgments}
The XMM-Newton project is supported by the Bundesministerium f\"ur Wirtschaft und 
Technologie/Deutsches Zentrum f\"ur Luft- und Raumfahrt (BMWI/DLR, FKZ 50 OX 0001)
and the Max-Planck Society. 
We acknowledge the use of archival \swift\ data and thank the \swift\ team for scheduling new observations.
The OGLE project has received funding from the European Research Council
under the European Community's Seventh Framework Programme
(FP7/2007-2013) / ERC grant agreement no. 246678 to AU.
We are grateful to S. Schmidl (Th\"uringer Landessternwarte Tautenburg)
for overlooking the GROND observation.
Part of the funding for GROND (both hardware as well as personnel) was 
generously granted from the Leibniz-Prize to Prof. G. Hasinger 
(DFG grant HA 1850/28-1). 
PM and GV acknowledge support from the BMWI/DLR grant FKZ 50 OR 1201 and FKZ 50 OR 1208, respectively.

\bibliographystyle{mn2e}
\bibliography{mnras_sxp265}

\end{document}